\documentclass{raa01}

\usepackage{graphicx,times}
\usepackage{natbib}
\bibpunct{(}{)}{;}{a}{}{,}
\usepackage{upgreek}

\begin{document}

   \title{Optical/IR studies of Be stars in NGC 6834 with emphasis on two specific stars}

\volnopage{{\bf 2014} Vol.\ {\bf 14} No. {\bf 9}, 1173--1192~~
 {\small doi: 10.1088/1674--4527/14/9/008}}
   \setcounter{page}{1173}

   \author{Blesson Mathew
               \inst{1,4}
   \and Watson P. Varricatt
               \inst{2}
   \and Annapurni Subramaniam
               \inst{3}
   \and N. M. Ashok
               \inst{4}
   \and  \\[1mm] D. P. K. Banerjee
               \inst{4}
          }

   \institute{Centre for Astrophysics \& Supercomputing, Swinburne University,
     Hawthorn VIC 3122, Australia; {\it blessonmathew@gmail.com}\\
         \and
             Joint Astronomy Centre, 660 N. Aohoku Pl., Hilo, HI{ }96720, USA \\
         \and
             Indian Institute of Astrophysics, Bangalore - 560 034, India \\
         \and
             Astronomy and Astrophysics Division, Physical Research
             Laboratory, Navrangapura, Ahmedabad - 380 009, Gujarat, India \\
\vs\no
   {\small Received 2014 February 12; accepted 2014 March 24}}

\abstract{  We present optical and infrared photometric and
spectroscopic studies of two Be stars in the {70--80-Myr-old open}
cluster NGC~6834. NGC~6834(1) has been reported as a binary from
speckle interferometric studies whereas NGC~6834(2) may possibly
be a $\gamma$ Cas-like variable. Infrared photometry and
spectroscopy from {the United Kingdom Infrared Telescope
(}UKIRT{)}, and optical data from various facilities are combined
with archival data to understand the nature of these candidates.
High signal-to-noise near-IR spectra obtained from UKIRT {have
}enable{d} us to study the optical depth effects in the hydrogen
emission lines {of} these stars. We have explored the spectral
classification scheme based on the intensity of emission lines in
{the }$H$ and $K$ bands and contrasted it with the conventional
classification based on the intensity of hydrogen and helium
absorption lines. This work also presents hitherto unavailable
{\it UBV} CCD photometry of NGC~6834, from which the evolutionary
state of the Be stars is identified. 
}

     \authorrunning{B. Mathew et al.}
     \titlerunning{Be Stars in NGC~6834}

     \maketitle

\section{INTRODUCTION}
\label{sect:intro}

Classical Be stars are non-supergiant B-type stars{ that} show, or
{have} shown{,} Balmer emission lines in their spectra at least
once in their lifetime (\citealt{Collins+1987};
\citealt{Porter+Rivinius+2003}). These, and other recombination
lines of hydrogen, are formed in a circumstellar disk that is
formed through the decretion mechanism (see
\citealt{Rivinius+etal+2013} for a recent review). Be stars rotate
close to the critical velocity (\citealt{Townsend+etal+2004}).
This, along with non-radial pulsations, assists the formation of
the dis{k} (also known as the `Be phenomenon{,}'
\citealt{Slettebak+1988}). Although recent theoretical models
{such as the} viscous decretion dis{k} model explain the formation
of the dis{k} once the mass is accumulated in the circumstellar
medium, the mode of mass transfer from the star to the dis{k} is
still an open issue. This problem can be resolved from the study
of a large sample of Be stars in diverse environments, {such as}
in the galactic field and in open clusters. The near-infrared
spectral region is ideally suited for studies of this kind since
the extinction is much lower (only {$\sim$}one-tenth that of
optical in the $K$ band) and `IR excess' starts showing up in the
{\it JHK} bands (\citealt{Dougherty+etal+1994}). From the
pioneering study of \cite{Gehrz+etal+1974}, it is understood that
the dis{k} contributes excess flux in the infrared, which appears
as IR excess over the stellar photospheric continuum. The IR
excess has been quantified by means of free-free and free-bound
emission from the electrons in the gaseous dis{k} without an
additional contribution from dust (\citealt{Ashok+etal+1984};
\citealt{Banerjee+etal+2001}). In addition to continuum emission,
the H{\sc i} recombination lines formed in the dis{k} fill up the
photospheric absorption profiles and appear as emission lines. The
H{\sc i} recombination lines are subjected to opacity effects,
which suggest that the line formation region in the dis{k} is
extremely dense, with an electron density of 10$^{10}$ --
10$^{14}$ cm$^{-3}$ (\citealt{Mathew+etal+2012b};
\citealt{Ashok+Banerjee+2000}). \cite{Clark+Steele+2000} were the
first to conduct a systematic study on the spectra of Be stars in
{the }$K$ band, which was followed by a similar study in {the }$H$
band by \cite{Steele+Clark+2001}. They proposed a scheme to
identify the spectral types of Be stars from the intensity of
emission lines in the $H$ and $K$ bands, which is a promising
method to classify highly reddened Be stars that are counterparts
of X-ray transient systems.

To gain a better understanding {of} the Be phenomenon it is
necessary to study the infrared spectra of Be stars in diverse
environments. All {of the }previous spectroscopic studies of Be
stars, especially in the infrared, {have} concentrated on {nearby
}field stars. The prime reason is that they are within a distance
of 1\,kpc, which makes them bright enough to be observed even with
1-m class telescopes. This paper presents the first of our efforts
to study the infrared spectra of Be stars in young ($<$100~Myr)
open clusters, at distances greater than 1\,kpc. In this work we
present infrared spectroscopy of two Be stars associated with the
cluster NGC~6834, located at a distance of 3.1\,kpc.

NGC~6834 is a young open cluster in Cygnus, belonging to the
Trumpler type~II~2~m ({II} - detached cluster with a little
central concentration, 2 - moderate range in brightness, m -
medium rich (50--100 stars), \citealt{Ruprecht+1966}). A range of
distance estimates are available for this cluster - from 2100\,pc
reported by \cite{Funfschilling+1967} to 2750\,pc from the studies
of \cite{Miller+etal+1996}. \cite{Paunzen+etal+2006} estimated a
distance of 1930$\pm$32~pc from CCD observations in{ the}
$g_1g_2y$ photometric system, which matches the earlier estimates
by \cite{Trumpler+1930}. Similarly, one can find divers{e} age
estimates; \cite{Miller+etal+1996} estimated the age of the
cluster to be around 50\,Myr, while \cite{Moffat+1972} assigned a
value of 80\,Myr. \cite{Paunzen+etal+2006} estimated a mean color
excess $E(B-V)$ = 0.70$\pm$0.05~mag, and
\cite{Jerzykiewicz+etal+2011} found a range in $E(B-V)$ values
between 0.61 and 0.82~mag. \cite{Jerzykiewicz+etal+2011}
identified 15 B-type variable stars in the cluster. They found
five cluster members showing H$\alpha$ emission, including a
$\gamma$ Cas-type variable and two $\lambda$ Eri-type variables.
\cite{Miller+etal+1996} found six Be stars in the cluster from
broadband photometry and observations using narrow-band filters
centered on the H$\alpha$ line and the adjacent continuum.
\cite{Mathew+etal+2008}, in their slitless spectroscopic survey to
identify emission-line stars in young Galactic open clusters,
identified four Be stars in this cluster. The Be stars showing
faint H$\alpha$ emission that barely fills up the absorption
profile and remains below the continuum level will not be detected
through slitless spectroscopy. Hence{,} there {could} be many more
Be stars in the cluster, as reported by \cite{Miller+etal+1996}
and \cite{Jerzykiewicz+etal+2011}. However, since th{e}se
detections were based on narrow-band H$\alpha$ photometry, the Be
status has to be confirmed through spectroscopy{ and until this}
is done we assume that there are only four Be stars in NGC~6834,
whose spectral analysis {has been} done and {whose} Be status {has
been} confirmed (\citealt{Mathew+Subramaniam+2011}).

\begin{figure}
\centering
\includegraphics[width=7cm]{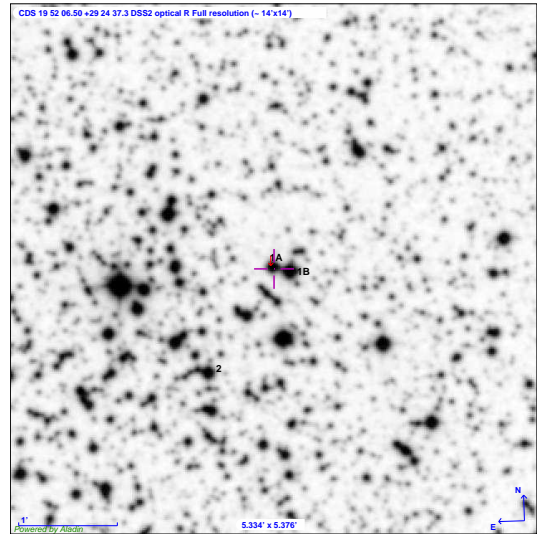}

\caption{\baselineskip 3.6mm The Be stars NGC~6834(1) and
NGC~6834(2) are shown in{ the} POSS~II plate,
  labeled 1A and 2{,} respectively. Also shown is the possible companion to
  NGC~6834(1), reported by \cite{Mason+etal+2002}, as 1B.\label{fig1}}
\end{figure}

The Be stars were cataloged as NGC~6834(1), NGC~6834(2),
NGC~6834(3) and NGC~6834(4) in \cite{Mathew+Subramaniam+2011},
while their corresponding numbers in WEBDA are 55, 67, 106 and 36
(based on the numbering scheme of \citealt{Funfschilling+1967}).
\cite{Jerzykiewicz+etal+2011} identified the stars 55 and 67
(named by them V6 and V2 respectively) {as} variables, with the
latter classified as a $\gamma$ Cas-like variable. Among the four
Be stars, NGC~6834(1) and NGC~6834(2) show{ a} high value of
H$\alpha$ equivalent width{ (EW)} ($\sim$ 40\,\AA), which is also
a reasonably high value among the sample of classical Be stars in
open clusters (\citealt{Mathew+Subramaniam+2011}). The intense
H$\alpha$ emission in these two candidates is supported from the
studies of \cite{Jerzykiewicz+etal+2011}, who quantified it by
means of the $\alpha$ index (H$\alpha$$_{\rm
line}$--H$\alpha$$_{\rm continuum}$). The Be stars NGC~6834(1) and
NGC~6834(2) are shown in Figure~\ref{fig1}. 

The motivations for the present study are summarized as follows:

\begin{itemize}
  \item[(1)] Previous infrared spectroscopic studies of Be stars {have}
concentrated on bright, nearby field Be stars. To the best of our
knowledge, this is the first effort to study Be stars {that} are
members of young ($<$100\,Myr), distant ($>$1\,kpc) open clusters.
More studies are planned in this direction, which will address the
role of{ the} cluster environment in the formation of {a }dis{k}
in Be stars. Most of the Be stars, which are members of open
clusters and{ were} identified in the survey by
\cite{Mathew+etal+2008, Mathew+Subramaniam+2011} are at a distance
greater than 3\,kpc; these sources can be studied in the infrared
using facilities with capacity similar to that of UKIRT. The IR
spectroscopy when combined with optical spectra and available
archival data (in a similar fashion as we have done in this work)
will provide insights about the nature of the circumstellar dis{k}
in Be stars.

\item[(2)]  A recombination analysis of the flux ratios of emission lines
belonging to Brackett and Pfund series of hydrogen, {as well as}
{a }comparison with the theoretical estimates, addresses the role
of optical depth effects in these systems. In{ the} future, we
plan to use optical spectra covering Balmer and higher order
Paschen lines, in conjunction with infrared spectra, which will
enable us to perform elaborate recombination line analysis over
optical and infrared bands.

\item[(3)] The stars NGC~6834(1) and NGC~6834(2) are selected for this study
due to the following reasons: (i) \cite{Mason+etal+2002}
identified a possible companion to NGC~6834(1) at a distance of
10.35$\arcsec$ from speckle interferometric studies (shown as 1B
in Fig.~\ref{fig1}){;} (ii) NGC~6834(2) has been found to exhibit
long-term and short-term variability{ and, b}ased on the
short-term variability and the presence of H$\alpha$ emission,
\cite{Jerzykiewicz+etal+2011} classified this star as a $\gamma$
Cas-like variable{; and}, (iii) {b}oth stars are found to exhibit
strong H$\alpha$ emission, which suggests the possible presence of
an extended circumstellar dis{k}.

\item[(4)] There is an uncertainty about the spectral types of these stars
(in Sect.~3.3 we re-assess their spectral types from optical and
infrared spectroscopy). {In addition}, a clear picture {of} the
evolutionary status of these stars can be obtained from the
optical color-magnitude diagram, {which is }constructed using the
newly obtained photometric data.

\end{itemize}

The paper is arranged as follows. Section~2 elaborates on the optical and infrared data used in this work. The results of this study are presented in Section~3. Based on the available observations, the nature of the two Be stars is discussed in Section~4.

\section{OBSERVATIONS AND ANALYSIS}
\label{sect:Obs} Near-infrared photometric and spectroscopic
observations of the target stars were obtained.  These data are
combined with new optical photometric and spectroscopic data{,}
and archival optical and IR data to gain a better understanding of
the objects under study.  The details of the observations and data
analysis are given below.

\subsection{Near-Infrared Spectroscopy}
Observations were carried out with the 3.8-m United Kingdom
Infrared Telescope{ (UKIRT)}, Hawaii, using the UKIRT
Imager-Spectrometer (UIST; \citealt{RamsayHowat+etal+2004}). A
4-pixel-wide slit (0.46$\arcsec$) was used, which gives a spectral
resolving power of 550 when observed with the $HK$ grism, and 1500
with the short-$J$ grism. The observations were performed by
alternating the object between two positions separated by
12$\arcsec$ along the slit. A black body unit mounted on the
instrument was used for flat fielding. Arc spectra obtained with
{an }argon lamp {were} used for wavelength calibration.
Preliminary data reduction {(}including flat fielding, subtracting
the dithered frames and coadding these subtracted pairs{)} {was}
carried out using the UKIRT pipeline ORACDR. The final extraction
of the spectra, wavelength calibration, ratioing and flux
calibration were done using the STARLINK packages FIGARO and
KAPPA, and the NOAO package Image Reduction and Analysis Facility
(IRAF).

To remove the telluric absorption bands from the object spectra,
BS~7793 (HD~194012; F8V) was observed immediately before the
object, at a similar airmass. We used the NASA Infrared Telescope
Facility (IRTF) spectrum of HD~219623
(\citealt{Rayner+etal+2009}){,} which has a similar spectral type,
taken from the NASA IRTF spectral library, to derive the  sky and
the instrument transmission function. The IRTF spectrum of
HD~219623, which has a higher spectral resolution{,} was smoothed
to match the resolution of the object spectra and {was }then
interpolated between 18050 and 19300\,\AA~where the sky
transmission is poor. The wavelength calibrated spectrum of
BS~7793 was then divided by the modified IRTF spectrum to remove
the slope of the stellar spectral energy distribution (SED) and
the photospheric absorption lines. The extracted, wavelength
calibrated object spectra were then divided by the ratioed
spectrum. The {\it HK} spectra were normalized at the central
wavelength of the $K$-band, and were then multiplied {by} the
2MASS $K$-band flux values to flux calibrate the object spectra.
The short-$J$ spectra were not flux calibrated since  the central
wavelength of the $J$ band filter falls outside the wavelength
range covered by the short-$J$ grism. The log of the observations
is given in Table~\ref{tab1}. 

\begin{table*}[h!!]
\centering

\begin{minipage}{80mm}

\caption{\baselineskip 3.6mm
Journal of infrared spectroscopic
observations from UKIRT. Observations were
  done on 2006--10--23.\label{tab1}}\end{minipage}

  \fns \tabcolsep 2mm
\begin{tabular}{lcccc}
\hline\noalign{\smallskip}
Object  &   Grism   &  UT    & Mean & Integration  \\
        &              &         & airmass &  time (s)\\
\noalign{\smallskip}\hline\noalign{\smallskip}
NGC~6834(1) &   {\it HK}    &  06:25:39 & 1.19  & 720 \\
NGC~6834(2) &   {\it HK}    &  06:42:35 & 1.24  & 480 \\
NGC~6834(1) &   short-$J$  &  07:08:10 & 1.35  & 1440 \\
NGC~6834(2) &   short-$J$  &  07:29:39 & 1.46  & 720 \\
\noalign{\smallskip}\hline
\end{tabular}

\vs\vs \small \centering
\begin{minipage}{104mm}
\caption{\baselineskip 3.6mm Journal of infrared photometric
observations from UKIRT. {\it JHK}
  photometry was performed on 2007--05--27 and $LM$ on 2007--06--19.\label{tab2}}
\end{minipage}

\fns\tabcolsep 3mm
\begin{tabular}{lcccc}
\hline\noalign{\smallskip}
Object  & Bandpass &  Integration & Mean & Magnitude\\
        &          &  time (s) & airmass & \\
\noalign{\smallskip}\hline\noalign{\smallskip}
NGC~6834(1) & $J$  & 135 & 1.05 & 11.923$\pm$0.012 \\
            & $H$  &  90 & 1.05 & 11.685$\pm$0.018 \\
            & $K$  &  90 & 1.06 & 11.482$\pm$0.015 \\
            & $L'$  &  80 & 1.06 & 10.98$\pm$0.08 \\
            & $M'$  & 307.2 & 1.07 & 11.16$\pm$0.48 \\
NGC~6834(2) & $J$  &  90 & 1.07 & 10.960$\pm$0.012 \\
            & $H$  &  45 & 1.06 & 10.697$\pm$0.018 \\
            & $K$  &  45 & 1.07 & 10.408$\pm$0.015 \\
            & $L'$  &  32 & 1.08 &  9.80$\pm$0.04 \\
            & $M'$  & 153.6 & 1.09 & 9.96$\pm$0.24 \\
\noalign{\smallskip}\hline
\end{tabular}
\end{table*}

\subsection{Infrared Photometry}
\subsubsection{{\it JHK} photometry}

The {\it JHK} photometry was performed with the UKIRT Fast-Track
Imager (UFTI; \citealt{Roche+etal+2003}) mounted on UKIRT.
Observations were carried out in photometric sky conditions (see
Table~\ref{tab2} for the observation log). The UKIRT faint
standard FS~149 (\citealt{Leggett+etal+2006}) was observed using
the 512$\times$512 sub-array, dithering the star to {five}
positions on the array, each with a separation of 10$\arcsec$. The
target fields were observed using the 1~k$\times$1~k full array. A
9-point dither pattern with offsets of 20$\arcsec$ from the
central position was adopted for the target observations.  A
single frame with UFTI gives a field of view of
93\arcsec$\times$93\arcsec on the sky at a plate scale of
0.091$\arcsec$ pixel$^{-1}$. With a 20$\arcsec$ dither pattern,
the final mosaic has a field of view of
2.2\arcmin$\times$2.2\arcmin. {A s}tandard star was observed just
before the target, at an airmass very close to that of the target.
Hence, the extinction corrections are far less than the
observational errors.

Two separate sets of target observations were obtained, each
around NGC~6834(1) and NGC~6834(2){,} respectively. Dark frames
were obtained before each set of target observations and they were
subtracted from the object frames. Flat fielding {was} done using
a flat field frame generated by median combining the dithered
object frames. The magnitudes of the Be stars were estimated
through{ the} standard reduction procedure in IRAF, and the values
are listed in Table~\ref{tab2}. 

Astrometric corrections were applied to our mosaics using the 2MASS
positions as reference.  2MASS positions of typically nine isolated
point sources were compared with the positions of these sources in
our mosaics and the required coordinate offsets were applied. The
J2000 coordinates (RA, Dec) derived for NGC~6834(1) and NGC~6834(2)
are (19:52:06.471, +29:24:37.70) and (19:52:09.528,
+29:23:34.14){,} respectively.

\subsubsection{$L'M'$ photometry}

The $L'$ and $M'$ observations were performed with{ the} UKIRT
$1-5~\upmu$m 
 imager spectrometer (UIST;
\citealt{RamsayHowat+etal+2004}) in photometric sky conditions.
The 0.12\arcsec\,pixel$^{-1}$ camera of UIST was used. For $L'$
observations, {a }1~k$\times$1~k full array was used, whereas {a
}512$\times$512 sub-array was used for $M'$. The object was
dithered to {four} positions on the array for $L'$, and to {eight}
positions on the array for $M'$. To counter the large sky
background, individual pairs within these 4- or 8-point jitters
were subtracted from each other and flat fielded with flat fields
generated by median combining the on-sky frames. These were then
combined to form the final mosaics, which cover a field of
1.95\arcmin$\times$1.65\arcmin~in $L'$ and
1.1\arcmin$\times$1.2\arcmin~in $M'$. The photometric standards
GL811.1 and GL748 were observed in {the }$L'$ and $M'$ bands{,}
respectively{,} at {an }airmass similar to that of the Be stars.
Aperture photometry was performed using STARLINK {\sc GAIA}; the
{derived }magnitudes of the Be stars are given in
Table~\ref{tab2}.

\subsubsection{{\it WISE} photometry}

The Wide field Infrared Survey Explorer ({\it WISE}) carried out
an all-sky survey from 3.4 to 22\,$\upmu$m with a sensitivity 100
times better than {\it IRAS}. The survey was done in four bands
centered at 3.4, 4.6, 12 and 22\,$\upmu$m, which are labeled $W1$,
$W2$, $W3$ and $W4$ respectively (Wright et al.\ 2010). The
angular resolution in the four bands {is} $6.1''$, $6.4''$,
$6.5''$, and $12\arcsec$ respectively. We queried for any
detection of our targets in the {\it WISE} all-sky catalog through
{the }IRSA database via the GATOR query engine. The search radius
was fixed at 6$\arcsec$, which is equivalent to the angular
resolution in {the }first {three} {\it WISE} bands. {\it WISE}
detected the sources in all {four} bands with definite errors and
we have used these magnitudes for the present study.

\subsection{Optical Spectroscopy}
\subsubsection{2.34-m Vainu Bappu Telescope}

Medium resolution spectra were obtained using the Optomechanics
Research (OMR) Spectrograph, mounted at the Cassegrain focus of
the 2.34-m Vainu Bappu Telescope (VBT), Kavalur, India (Prabhu et
al.\ 1998). The observations were performed with a 1200\,l/mm
grating, which provides an effective resolution of 4\,\AA~in the
region of interest (around H$\alpha$ profile). FeNe lamp spectra
were taken after the object for wavelength calibration. Dome flats
were taken to correct the object frames for pixel-to-pixel
efficiency. The extracted spectra were bias subtracted,
flat-fielded and wavelength calibrated in the standard way using
the software packages available in IRAF. The log of the
observations is given
in Table~\ref{tab3}. 

\begin{table*}[h]
\centering \begin{minipage}{80mm}\caption{Journal of Optical
Spectroscopic Observations\label{tab3}}\end{minipage}

\fns\tabcolsep 1mm
\begin{tabular}{lcccccc}
\hline\noalign{\smallskip}
Facility & Be star & Date    &  Integration & H$\alpha$ & He{\sc i} & He{\sc i}\\
         &         &         &   time (s) & EW {(\AA)}& 5876 EW {(\AA)}& 6678 EW {(\AA)}\\
\noalign{\smallskip}\hline\noalign{\smallskip}
VBT & NGC~6834(1) & 2007--11--14 & 2700 & --37.3 & --   &  -- \\
    & NGC~6834(2) & 2007--11--15 & 2700 & --34.7 & --   & 0.17\\
HCT & NGC~6834(1) & 2005--10--07 & 900  & --38.9 & 0.58 & 0.36\\
    &             & 2007--07--05 & 900  & --37.5 & 0.52 & 0.30\\
    & NGC~6834(2) & 2005--10--07 & 600  & --42.5 & 0.32 & 0.24\\
    &             & 2007--07--06 & 600  & --41.7 & 0.37 & 0.23\\
\noalign{\smallskip}\hline
\end{tabular}
\end{table*}

\subsubsection{2-m Himalayan Chandra Telescope}

The optical spectra of the Be stars in the wavelength range
5500--9000\,\AA\ were obtained using the Himalayan Faint Object
Spectrometer and Camera (HFOSC), mounted on{ the} 2.1-m Himalayan
Chandra Telescope (HCT). The spectra {have} a resolution of
7\,\AA~around the H$\alpha$ spectral region.
The data reduction follows the method outlined in
\cite{Mathew+Subramaniam+2011}. The log of the observations is
given in Table~\ref{tab3}. We have {two} sets of spectra for each
Be star - one before (2005--10--07) and another after
(2007--07--06) the UKIRT observation. This allows us to check
whether the source shows spectral variability in conjunction with
the infrared spectra.

\subsection{Optical Photometry}

Optical photometry of NGC~6834 was performed on 2003--09--19 with
HFOSC mounted on HCT. The cluster was observed in {\it UBV} with
three different exposure times, (3\,s, 10\,s, 60\,s) in $V$,
(5\,s, 60\,s, 180\,s) in $B$ and (30\,s, 180\,s, 600\,s) in $U$.
These frames with variable exposures were needed to do the
photometry of bright and faint cluster members. The nights were
photometric, and the Landolt standards were observed for
photometric calibrations. The zero point errors are 0.010, 0.013
and 0.018~mag 
 in $V$, $B$ and $U${,} respectively. The data reduction and
calibration procedure were the same as those elaborated in
\cite{Subramaniam+Bhatt+2007}. The magnitudes derived are given in
Table~\ref{tab5}.


\section{RESULTS}

\subsection{Revised Estimates of Cluster Parameters from New Photometry}

There is a scarcity of photometric data for this cluster, as seen
from the WEBDA database ({\it www.univie. ac.at/webda/}).

Section 1 shows that a range of values {is} available for cluster
parameters{;} hence{,} deep {\it UBV} CCD photometry which
includes the faint stars is needed to better constrain these
values. A precise estimation of age and distance 
for the cluster is important to constrain the evolutionary phase
of the Be stars.

\begin{figure}
\centering
\includegraphics[width=7.2cm]{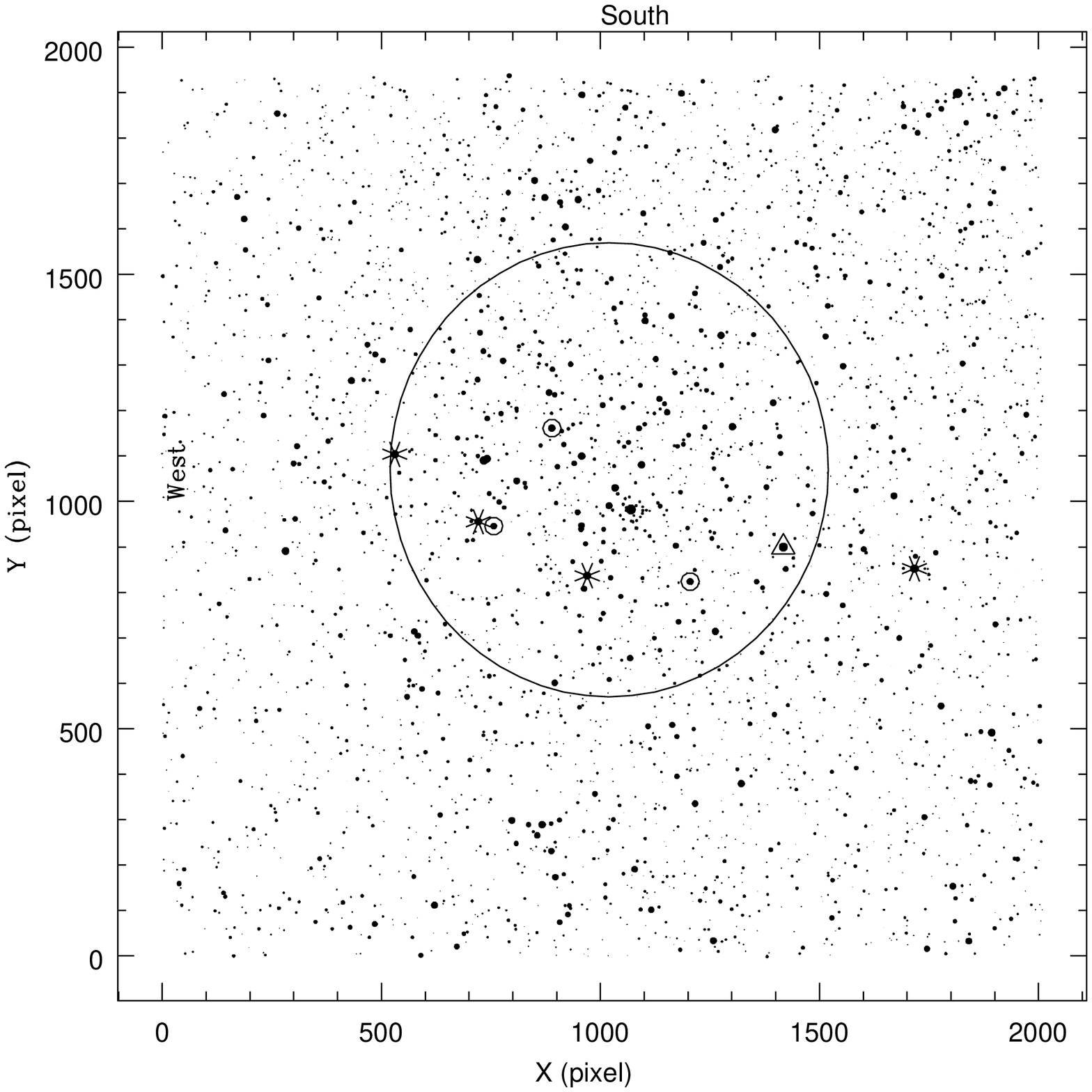}

\vspace{-3mm}

\caption{\baselineskip 3.6mm $X$-$Y$ plot of the cluster NGC~6834
is shown with the locations of the Be stars (encircled), giants
(in asterisks) and YSG (in an open triangle)
indicated.\label{fig2}}
\centering
\includegraphics[width=7.2cm]{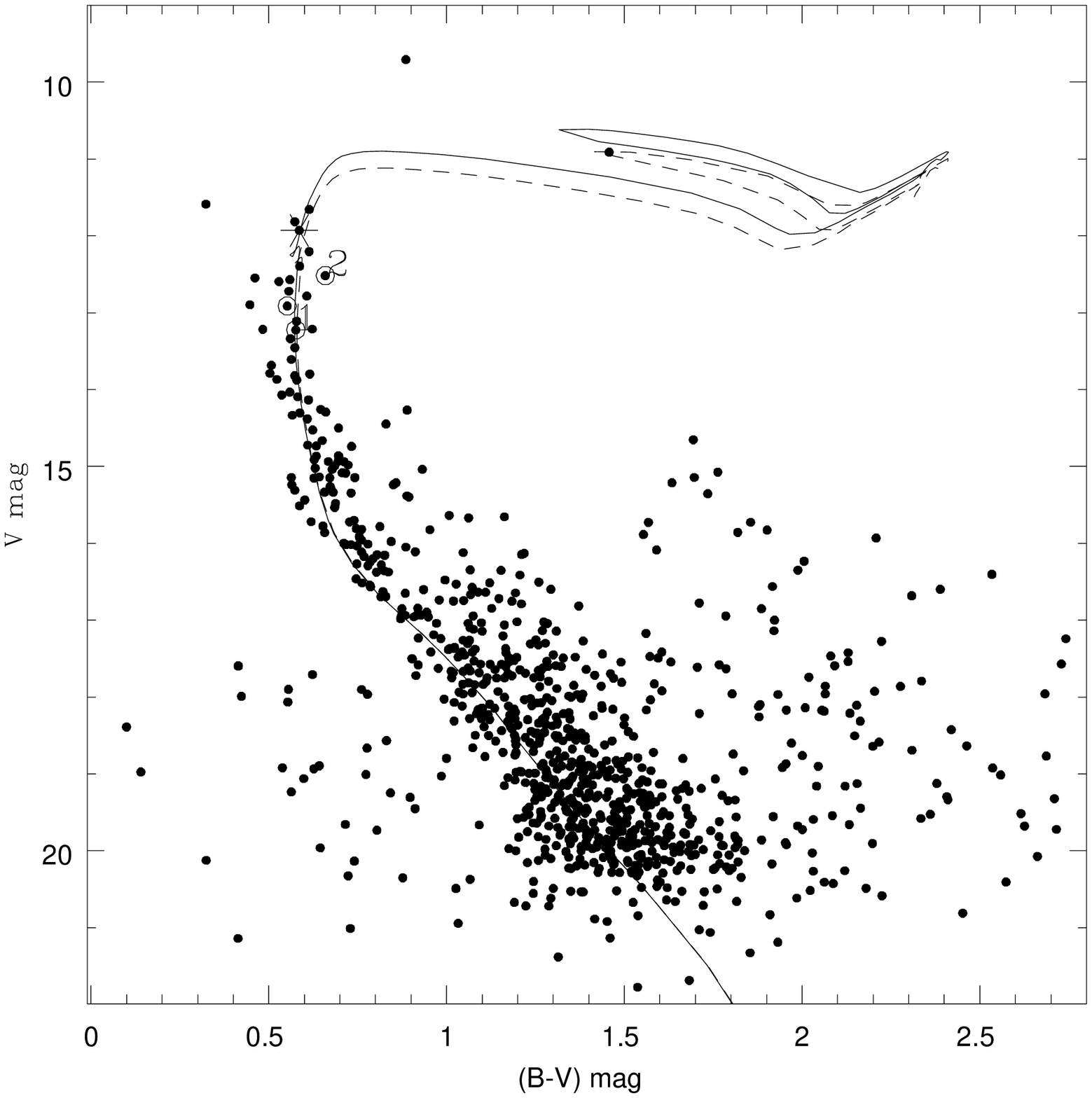}

\vspace{-3mm} \caption{\baselineskip 3.6mm The color-magnitude
diagram of NGC~6834. The Be stars identified are circled;
NGC~6834(1) and NGC~6834(2) are labeled as ``1'' and ``2''
respectively. The suspected binary companion to NGC~6834(1) is
shown with an asterisk. The solid and dashed lines are the
isochrones for ages 70 and 80\,Myr, respectively.\label{fig3}}
\end{figure}

We obtained {\it UBV} CCD photometry of the cluster and
re-estimated the cluster parameters. The $XY$ plot of the cluster
region is shown in Figure~\ref{fig2}{,} 
 with the Be
stars and the evolved stars shown with separate symbols. A circle
with a radius of 200\arcsec, which is the cluster radius estimated
from the radius density profile, is plotted {in}
Figure~\ref{fig2}. We derived a reddening value of $E(B-V)$ =
0.73, which closely agrees with the earlier estimates by
\cite{Moffat+1972} and \cite{Paunzen+etal+2006}, who found values
of 0.72 and 0.70{,} respectively. The distance modulus is
estimated to be 14.8{,} from which a distance of 3100\,pc is
derived. This places the cluster at a higher distance than the
previous estimates of 2750\,pc by \cite{Miller+etal+1996} and
1930\,pc by \cite{Paunzen+etal+2006}. \cite{Moffat+1972} suggested
that the limiting magnitude of the photometry should be low enough
to include the faint stars, which are needed to do a better {zero
age main sequence} fit to provide {an} accurate distance
measurement.
The superior nature of our photometry is relevant in this context{
because it} revises the distance of NGC~6834 to 3100\,pc. The age
is estimated from the isochrones of Marigo et al.\ (2008), which
indicates that most of the evolved candidates lie in the age range
of 70--80\,Myr. This new age measurement matches that of
\cite{Moffat+1972}, who found the cluster to be 80\,Myr old.

We have identified four giants associated with the cluster, three
within and one outside the cluster radius (see Fig.~\ref{fig2}).
Interestingly, one of these giants (shown {as an} asterisk in
Fig.~\ref{fig2} that is located {right beside a circled} Be star)
is identified as a companion to the Be star NGC~6834(1) from
speckle interferometric studies (\citealt{Mason+etal+2002}). {In
addition, a} yellow supergiant (YSG) is identified, which can be
clearly distinguished in the {color magnitude diagram (}CMD{)}.
 The YSG is near the blue
loop, and is close to becoming a Cepheid. They were not found to be
variables by \cite{Jerzykiewicz+etal+2011}. The Be stars in the
cluster are circled in Figure~\ref{fig2}. Photometry is not
available for NGC~6834(3), so it is not displayed.

The Be stars NGC~6834(1) and NGC~6834(2) are close to the evolved
part of the main sequence (Fig.~\ref{fig3}) 
 and they
have $V$ magnitudes of 13.224$\pm$0.006 and 12.518$\pm$0.003{,}
respectively. Assuming a standard extinction law and using the
revised cluster parameters ($E(B-V)$ = 0.73, $d = 3100$\,pc), the
spectral types of NGC~6834(1) and NGC~6834(2) are estimated to be
B3-5 and B2-3{,} respectively. Even though the Be stars are close
to the evolved part of the main sequence{,} they are {well
}separated from the giant star candidates identified in the
cluster. Hence, we assume them to be main sequence stars.
{E}stimates of their spectral types will be revisited in Section
3.3.

\subsection{Analysis of {\it JHK} Spectra}
\subsubsection{Description of UKIRT {\it JHK} spectra}

\begin{figure}
\centering
\includegraphics[width=8.2cm]{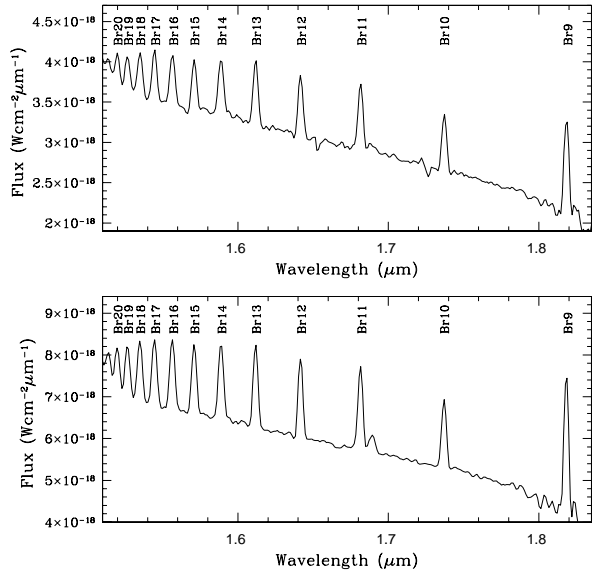}

\vspace{-2mm} \caption{\baselineskip 3.6mm Flux-calibrated
$H$-band spectra of NGC~6834(1) and NGC~6834(2) are shown in the
upper and lower panels respectively, with the prominent emission
lines labeled.\label{fig4}}
\end{figure}

\begin{figure}
\centering
\includegraphics[width=7.6cm]{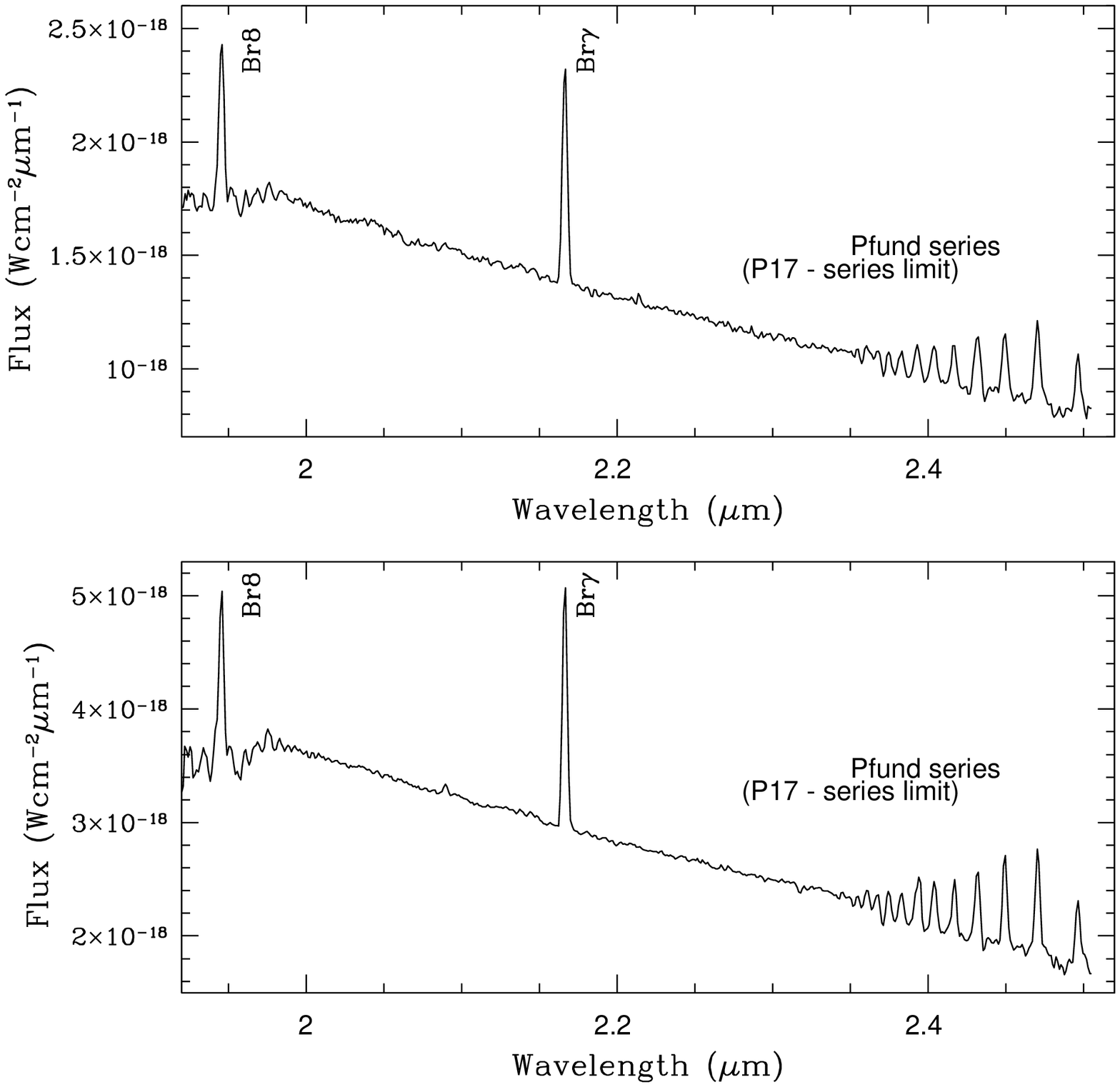}

\vspace{-3mm} \caption{\baselineskip 3.6mm Flux-calibrated
$K$-band spectra of NGC~6834(1) and NGC~6834(2) are shown in the
upper and lower panels respectively, with the prominent emission
lines labeled. \label{fig5}}
\end{figure}

Our {\it HK} spectra span a range of 1.5 -- 2.5\,$\upmu$m,
covering some of the prominent hydrogen recombination emission
lines belonging to the Brackett and Pfund series. The spectra of
the two Be stars are shown separately for clarity and to avoid the
region of poor atmospheric transmission. The spectra shown in
Figure~\ref{fig4} 
 cover the Brackett series from Br9
(1.8181\,$\upmu$m) to the series limit. The remaining emission
lines belonging to{ the} Brackett series (Br$\gamma$ and Br8) and
the higher order Pfund lines, from Pf17 to the series limit, are
shown in Figure~\ref{fig5}. 
 {The }Fe{\sc ii} 2.089~$\upmu$m
line is present in emission in the spectra of NGC~6834(2)
(Fig.~\ref{fig5}). Of the Be stars studied by
\cite{Granada+etal+2010}, only EW Lac showed {an }Fe{\sc ii}
2.089-$\upmu$m line in {its} spectra, which was identified as a
characteristic of moderately warm and dense environments. The
measured values of the {EW} and line flux of the Brackett and
Pfund series emission lines in the spectra of each of the Be stars
are shown in Table~\ref{tab4}. 
  The flux ratios of
these lines are used to study the optical depth effects in Be
stars, which is explained in the next section.


\begin{figure}
\centering
\includegraphics[width=7.3cm]{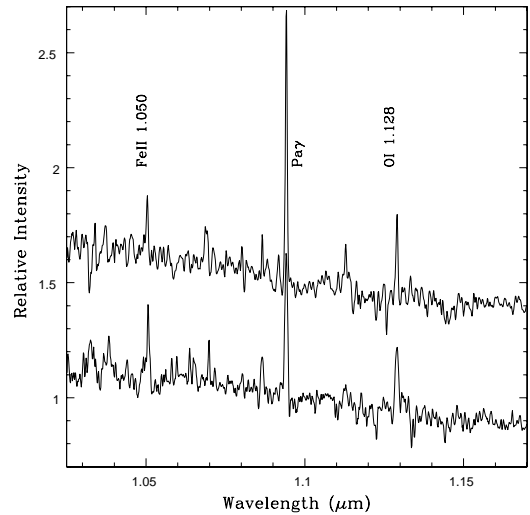}

\vspace{-3mm} \caption{\baselineskip 3.6mm The $J$-band spectra of
NGC~6834(1) ({\it lower}) and NGC~6834(2) ({\it upper}) shown with
a positive offset for clarity.\label{fig6}}
\end{figure}

The spectra obtained using the short-$J$ grism are shown in
Figure~\ref{fig6}. 
 The signal-to-noise ratio{s}
of the spectra are quite low, so we have smoothed {them}
for display purpose{s}. Of the hydrogen recombination lines, only
Pa$\gamma$ (1.0938~$\upmu$m) is visible in 
these spectra. Due to the small wavelength coverage of this grism,
Pa$\beta$ was not observed. The other interesting emission lines
in the spectra are due to O{\sc i} 1.1287\,$\upmu$m and Fe{\sc ii}
1.050\,$\upmu$m, which are labeled in Figure~\ref{fig6}. The
spectra were not flux calibrated, so we will not be including
these lines in further analysis. The presentation of these spectra
highlight{s} the absence of{ the} He{\sc i} 1.083\,$\upmu$m
emission line in both the Be stars, which can be corroborated with
the $HK$ spectra during the discussion on the spectral type of the
Be stars.

\begin{table*}

\centering \begin{minipage}{98mm}\caption{{EW} and Line Flux
Measurements from the Infrared Spectra\label{tab4}}\end{minipage}

\fns
\begin{tabular}{lccccc}
\hline\noalign{\smallskip}
  & & \multicolumn{2}{c}{NGC~6834(1)} & \multicolumn{2}{c}{NGC~6834(2)}\\
Line & $\lambda$ & EW & Line flux & EW & Line flux \\
     &  ($\upmu$m)   & (\AA) & (Wm$^{-2}$) & (\AA) & (Wm$^{-2}$) \\
\noalign{\smallskip}\hline\noalign{\smallskip}
Pf17 & 2.4953 & --12.3 & 9.967$\times10^{-18}$ & --11.7 & 2.081$\times10^{-17}$ \\
Pf18 & 2.4700 & --17.7 & 1.523$\times10^{-17}$ & --15.4 & 2.963$\times10^{-17}$ \\
Pf19 & 2.4490 & --12.1 & 1.084$\times10^{-17}$ & --13.8 & 2.676$\times10^{-17}$ \\
Pf20 & 2.4314 & --13.9 & 1.238$\times10^{-17}$ & --11.9 & 2.350$\times10^{-17}$ \\
Pf21 & 2.4164 & --7.3  & 6.831$\times10^{-18}$ & --8.3  & 1.678$\times10^{-17}$ \\
Pf22 & 2.4035 & --7.9  & 7.517$\times10^{-18}$ & --8.3  & 1.703$\times10^{-17}$ \\
Pf23 & 2.3925 & --7.1  & 6.762$\times10^{-18}$ & --8.9  & 1.867$\times10^{-17}$ \\
Pf24 & 2.3828 & --4.5  & 4.380$\times10^{-18}$ & --4.6  & 9.872$\times10^{-18}$ \\
Pf25 & 2.3744 & --3.9  & 3.808$\times10^{-18}$ & --4.3  & 9.155$\times10^{-18}$ \\
Br7  & 2.1661 & --28.1 & 3.861$\times10^{-17}$ & --26.3 & 7.797$\times10^{-17}$ \\
Br8  & 1.9451 & --15.7 & 2.730$\times10^{-17}$ & --14.5 & 5.251$\times10^{-17}$ \\
Br9  & 1.8181 & --21.8 & 4.591$\times10^{-17}$ & --24.8 & 1.089$\times10^{-16}$ \\
Br10 & 1.7367 & --10.2 & 2.688$\times10^{-17}$ & --11.6 & 6.174$\times10^{-17}$ \\
Br11 & 1.6811 & --10.2 & 2.996$\times10^{-17}$ & --12.7 & 7.358$\times10^{-17}$ \\
Br12 & 1.6412 & --10.0 & 3.057$\times10^{-17}$ & --11.7 & 7.083$\times10^{-17}$ \\
Br13 & 1.6114 & --9.5  & 3.064$\times10^{-17}$ & --11.7 & 7.380$\times10^{-17}$ \\
Br14 & 1.5885 & --7.9  & 2.697$\times10^{-17}$ & --10.7 & 6.999$\times10^{-17}$ \\
Br15 & 1.5705 & --6.8  & 2.310$\times10^{-17}$ & --9.6  & 6.381$\times10^{-17}$ \\
Br16 & 1.5561 & --6.8  & 2.387$\times10^{-17}$ & --10.1 & 6.778$\times10^{-17}$ \\
Br17 & 1.5443 & --6.9  & 2.451$\times10^{-17}$ & --9.3  & 6.308$\times10^{-17}$ \\
Br18 & 1.5346 & --5.3  & 1.910$\times10^{-17}$ & --7.7  & 5.340$\times10^{-17}$ \\
\noalign{\smallskip}\hline
\end{tabular}
\end{table*}

\subsubsection{Case~B analysis of the hydrogen emission lines}

From the conventional definition put forth by Baker \& Menzel
(1938), the Case~B condition {indicates} that the Lyman lines are
considered to be optically thick while all the other lines are
assumed to be optically thin. \cite{Hummer+Storey+1987} estimated
the emission strength of H{\sc i} recombination lines under Case~B
conditions for a range of temperatures and densities (especially
at higher densities, {\it viz.} $n_{\rm e} > 10^{10}$\,cm$^{-3}$)
where the collision effects are important. In the case of Be
stars, H{\sc i} recombination lines are formed in dense ($n_{\rm
e} = 10^{10} - 10^{14}$~cm$^{-3}$) regions of the circumstellar
dis{k} and{,} hence{,} we expect these lines to be optically thick
(\citealt{Ashok+Banerjee+2000}; \citealt{Steele+Clark+2001}).
\cite{Lenorzer+etal+2002} suggested{ using an} Hu14/Br$\gamma$
versus Hu14/Pf$\gamma$ flux ratio diagram as the diagnostic to
separate stars{,} which showed opacity effects from those for
which H{\sc i} emission strength followed Case~B values. A similar
method was followed by \cite{Granada+etal+2010}, who identified
the deviation of the emission strength of H{\sc i} emission lines
belonging to Pfund and Humphrey series from optically thin Case~B
values. However, a comparison between the flux values of the H{\sc
i} recombination lines (obtained from the infrared spectra) with
the theoretical estimates by \cite{Storey+Hummer+1995} will give a
quantitative picture of the optical depth effects in Be stars.
\cite{Mathew+etal+2012b} did {this }in the case of X Persei and
found the observed flux values of the H{\sc i} recombination lines
belonging to Paschen and Brackett series to be higher than the
theoretical values, thereby suggesting that these lines are
optically thick. In this work, we extend the analysis to longer
wavelengths, including{ the} Pfund series of H{\sc i} emission
lines.

\begin{figure}
\centering
\includegraphics[width=7.2cm]{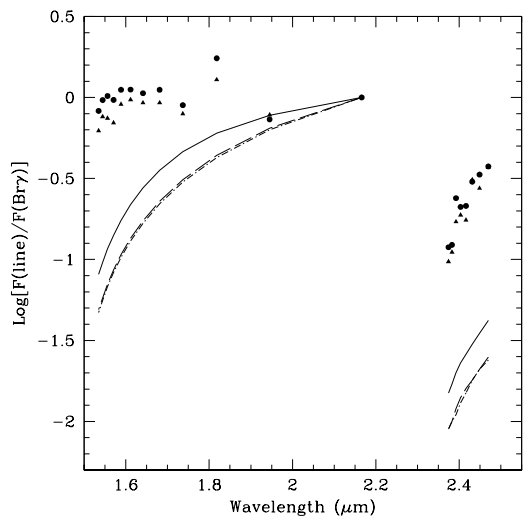}

\vspace{-2mm} \caption{\baselineskip 3.6mm Case~B recombination
analysis of the Brackett and Pfund series lines of hydrogen for
NGC~6834(1) and NGC~6834(2). The line fluxes are normalized with
respect to that of Br$\gamma$. The lines shown are Brackett 18 --
Pfund 25 ($1.5346 - 2.3744$\,$\upmu$m). The flux values for
NGC~6834(1) and NGC~6834(2) are shown in filled triangles and
filled circles{,} respectively. The recombination analysis is
carried out for $T_{\rm e} = 10^4$~K. The continuous, dashed and
dot-dashed lines show the Case~B values for $n_{\rm e}$ =
10$^{10}$, 10$^{12}$ and 10$^{14}$~cm$^{-3}$
respectively.\label{fig7}}
\end{figure}

The flux calibrated spectra of the two Be stars are dereddened for
a cluster{ with} $E(B-V)$ of 0.73, using the task DEREDDEN {in}
IRAF. The line flux values are measured for all the H{\sc i}
recombination lines belonging to{ the} Brackett and Pfund series
present in our spectra. The flux values of the Brackett lines Br8
(1.9451~$\upmu$m) and Br9 (1.8181~$\upmu$m) are susceptible to
large errors {because} they are located in regions of poor
atmospheric transmission. The flux ratios are calculated with
respect to Br$\gamma$ and are shown in Figure~\ref{fig7}.
 The Case~B emissivity values corresponding to $T_{\rm e} =
10^4$~K and $n_{\rm e} = 10^{10}$, 10$^{12}${ and}
10$^{14}$~cm$^{-3}$ are obtained from \cite{Storey+Hummer+1995},
and are shown using continuous lines in Figure~\ref{fig7}. We have
assumed the dis{k} to be isothermal with a mean temperature of
10$^4$\,K. The range of density values used for this study
{represents} typical values of density prevailing in Be star
dis{k}s (e.g. Waters 1986; \citealt{Silaj+etal+2010}). The
observed flux ratios of all the H{\sc i} lines belonging to
Brackett and the Pfund series (Br$\gamma$ -- Br18 and Pf17 --
Pf25) with respect to Br$\gamma$ are higher than the theoretical
estimates{, which} implies that these lines are optically thick.

\subsection{Re-estimation of the Spectral Type of the Be Stars}

There is ambiguity about the accuracy of spectral types of Be
stars. Even the spectral types of well-studied field Be stars from
the catalog of \cite{Jaschek+Egret+1982} have been re-estimated a
number of times using spectra {with} better sensitivity and
resolution because the photospheric line and continuum emission
can get modified by the emission from the circumstellar dis{k}.
From the revised {\it UBV} CCD photometry, we have estimated the
spectral types of NGC~6834(1) and NGC~6834(2) to be B3-5V and
B2-3V{,} respectively (as shown in Section 3.1). Classical Be
stars are known to have circumstellar dis{k}s, which redden the
stellar light, thereby biasing the estimates towards late spectral
types (\citealt{Slettebak+1982}). A conventional way to estimate
the spectral type from optical spectra is by matching the
intensity of the prominent higher order H{\sc i} (from H$\delta$
to H8) and He{\sc i} absorption lines (4471, 4026, 4143~\AA) {from
a spectral} library. The method is elaborated in
\cite{Mathew+Subramaniam+2011}{.} {A}dopting this technique we
estimate spectral types as B5V and B1V for NGC~6834(1) and
NGC~6834(2){,} respectively. The luminosity classification {is
consistent} with the absence of{ the} Mg{\sc ii} 4481\,\AA\
absorption line in the spectra of both the Be stars, which are
normally seen in the spectra of giants. However, there {may} still
be pitfalls in the classification scheme since hydrogen and helium
absorption lines can be filled-in by emission from the dis{k}
(even though we did a better job by using the lines least affected
by contamination for classification). Hence, we consider an
alternative scheme to do the spectral classification.

The present effort is to re-estimate the spectral types of
NGC~6834(1) and NGC~6834(2) from the relative intensities of the
emission lines in our 8300--8900\,\AA~optical spectra and
near-infrared {\it JHK} spectra. We use the studies by
\cite{Andrillat+etal+1988} and \cite{Clark+Steele+2000} as
guidance for classification{ in this study}.

\subsubsection{From optical spectra}

\begin{figure}

 \centering
\includegraphics[width=8.cm]{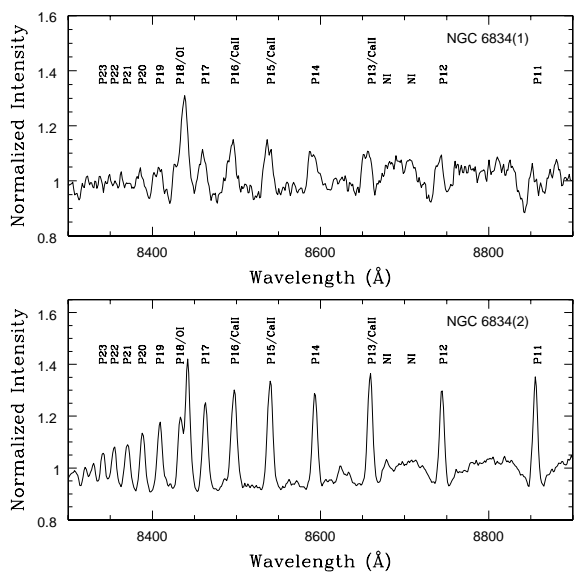}

\vspace{-3mm} \caption{\baselineskip 3.6mm The optical spectra of
the Be stars NGC~6834(1) and NGC~6834(2) in the wavelength region
8300--8900\,\AA.\label{fig8}}
\end{figure}

The 8300--8900\,\AA~optical spectra have been conventionally used
to estimate the luminosity class of Be stars (e.g.
\citealt{Andrillat+etal+1988}; Torrejon et al.\ 2010). The optical
spectra of both Be stars in this wavelength regime are presented
in Figure~\ref{fig8}, 
 and the prominent emission lines
are labeled. The major emission lines are due to:

\begin{itemize}
  \item[(1)] The neutral
hydrogen lines belonging to the Paschen series -- P11(8862),
P12(8750), P13(8665), P14(8598), P15(8545), P16(8502), P17(8467),
P18(8437), P19(8413), P20(8392), P21(8374), P22(8359),
P23(8345){;}
\item[(2)] Ca{\sc ii} triplet -- $\lambda$8498, 8542, 8662{;}
\item[(3)] O{\sc i} 7772\,\AA\ (Multiplet M1; 7772-7774-7775;
\citealt{Moore+1945}), 8446\,\AA\ (Multiplet M4;
\citealt{Moore+1945}){;}
\item[(4)] N{\sc i} composite $\lambda$8686-8683-8680 and $\lambda$8719-8712-8703{;}
\item[(5)] Fe{\sc ii} $\lambda$7712.
\end{itemize}

The inclusion of O{\sc i} $\lambda$7772 and Fe{\sc ii}
$\lambda$7712 in Figure 8 will affect the clarity of{ lines in}
the Paschen series, so {they} are not shown.

All the Paschen lines are present in {the }emission{s}
{from} both stars, with the lines being more intense in
NGC~6834(2) when compared to those in NGC~6834(1). The Paschen lines
P13, P15 and P16 are stronger than the other lines in the series,
which shows that they{ are} blended with the Ca{\sc ii} triplet. O{\sc i}
$\lambda$7772 and $\lambda$8446 are seen in emission in both the
stars. Fe{\sc ii} $\lambda$7712 is present in emission in
NGC~6834(2) whereas the emission is quite faint in NGC~6834(1). The
lower order Paschen lines, especially P11 and P12{,} are
embedded in the broad stellar absorption profile. Another notable
point is the presence of the whole Paschen series of emission lines,
down to the series limit, in the case of NGC~6834(2) {but} the
series is more or less truncated by P20 in NGC~6834(1).

The presence of Paschen emission lines suggest{s} that these Be
stars have spectral types earlier than B3 since these lines show
either fill-in features or absorption for later spectral types
(\citealt{Andrillat+etal+1988}). Further{more},{ the} Ca{\sc ii}
triplet and Fe{\sc ii} $\lambda$7712 lines are present in emission
in Be stars earlier than B2.5 (\citealt{Andrillat+etal+1988}).
However, it is better to measure the EW{s} and full-width at half
maximum (FWHM) of these stars, and {then }compare them with the
template spectra given by \cite{Andrillat+etal+1988}. The spectra
of NGC~6834(1) closely resemble that of the B0.5IV Be star HD~5394
in terms of the {EW}s of the Paschen lines and the truncation of{
the} Paschen series at P20. The EW values of the Paschen lines of
NGC~6834(2) agree with those of the Be star HD~148184 (for
{example}, EW(P14)$\sim$2.8), which is reported {to be} a B1.5Ve
star. These new spectral types based on the strength of emission
lines {have} to be contrasted with the earlier estimates, based on
the strength of He{\sc i} and H{\sc i} absorption lines. There is
{little} difference between the spectral type of NGC~6834(2)
estimated from both methods - B1 from the absorption lines, and
B1.5 from the emission lines. However, NGC~6834(1) has {been}
modified to an early-type star (B0.5) whereas the earlier
estimates indicated it {is a} B5. From a visual examination
(spectra not shown here), the hydrogen absorption lines from
H$\gamma$ to H8 seem to be intense in NGC~6834(1) (when compared
to NGC~6834(2)) whereas He{\sc i}  4471, 4026, 4143\,\AA\ lines
are intense in NGC~6834(2). Hence, NGC~6834(1) should be {a} later
spectral type when compared with NGC~6834(2), which also agrees
with the spectral estimates from photometry. This suggests that a
classification scheme based on Paschen emission lines is only
suggestive when the optical spectra in{ the} visual region are not
available or when the hydrogen and helium absorption lines are
veiled, particularly in the case of pre-main sequence stars
(\citealt{Mathew+etal+2012a}). Hence, at this point, NGC~6834(1)
stands as{ a} B5V star while NGC~6834(2) {is} modified to
B1-B1.5V. The next step is to see if the estimates agree with the
emission lines seen in the infrared spectra.

\subsubsection{From infrared spectra}
\cite{Clark+Steele+2000} introduced a spectral classification scheme in the
infrared, wherein the Be stars are classified into five groups based on
the presence and the relative strength of Br$\gamma$, He{\sc i} and
Mg{\sc ii} features in the $K$-band spectra. This is particularly useful in the
spectral classification of highly-reddened counterparts of X-ray transient
systems, which are subjected to large extinction in the optical.

Our sample of Be stars belong{s} to Group 5 of
\cite{Clark+Steele+2000} by virtue of the presence of Br$\gamma$
emission and the absence of He{\sc i} 1.083\,$\upmu$m,
2.058\,$\upmu$m and Mg{\sc ii} 2.138/2.144\,$\upmu$m spectral
features (Fig.\,5 \& 6). Further{more}, \cite{Clark+Steele+2000}
found that Group 5 Be stars belong to B5 or later spectral types.
Br$\gamma$ emission lines are intense in our sample of Be stars,
with the {EW} values of NGC~6834(1) and NGC~6834(2) being 28 and
26\,\AA\ respectively. Interestingly, {the} strong emission in
Br$\gamma$ is not seen in any of the Be stars presented in
\cite{Clark+Steele+2000}.

This opens up a new riddle. For NGC~6834(1), the optical and
infrared data suggest a B5 spectral type. However, NGC~6834(2) has
been identified to be later than B5 from the infrared
classification scheme, whereas we derive a spectra{l} type of
B1-1.5 from the optical spectra. NGC~6834(2) has been reported to
be a variable (\citealt{Jerzykiewicz+etal+2011}). It can be seen
from Table~\ref{tab3} that{ lines from} He{\sc i} 6678\,\AA\ show
variability between two epochs of observations. Therefore, we
cannot rule out the possibility of a varying He{\sc i}
2.058\,$\upmu$m profile in NGC~6834(2). The possibility of a
companion influencing the spectral line features should also be
considered since NGC~6834(2) has been suggested as a $\gamma$
Cas-like variable. For the present purpose, we conclude the
spectral type of Be stars NGC~6834(1) and NGC~6834(2) to be B5V
and B1-1.5V{,} respectively.

We would like to point out the uncertainty of a classification
scheme based on emission lines since the region of line
formation in the dis{k} is highly active, thereby inducing variability
in spectral lines. {In addition}, the group
classification scheme of \cite{Clark+Steele+2000} needs a revision
by including {a larger} sample of Be stars since we found that the
Br$\gamma$ emission lines in NGC~6834(1) and NGC~6834(2) are{ more} intense
than any of the stars discussed in \cite{Clark+Steele+2000}.

\begin{table*}
\centering \caption{\baselineskip 3.6mm
 Available magnitudes and
colors of the candidate Be stars. Also given are the magnitudes of
NGC~6834(1B), which is the proposed companion of the Be star
NGC~6834(1).\label{tab5}}

\vspace{-3mm} \fns\tabcolsep 5mm

\begin{tabular}{lcccc} \hline\noalign{\smallskip}
Reference & Band & NGC~6834(1) & NGC~6834(2) & NGC~6834(1B)\\
\noalign{\smallskip}\hline\noalign{\smallskip}
This work & $V$ & 13.224$\pm$0.006 & 12.518$\pm$0.003 & 11.929$\pm$0.005\\
          & $(B-V)$ & 0.576$\pm$0.006 & 0.659$\pm$0.003 & 0.586$\pm$0.010\\
HMUBV & $V$ & 13.20 & -- & --\\
      & $(B-V)$ & 0.59 & -- & --\\
      & $(U-B)$ & --0.15 & -- & --\\
P06   & $V$ & 13.22 & 12.47 & 11.95\\
      & $(B-V)$ & 0.666 & 0.797 & 0.693\\
H61   & $V$ & 13.210 & -- & 11.940 \\
      & $(B-V)$ & 0.590 & -- & 0.610\\
      & $(U-B)$ & --0.150 & -- & 0.040\\
JKP11 & $V$ & 13.117 & 12.463 & --\\
      & $(V-I_C)$ & 0.841 & 0.950 & --\\
KW97  & $V$ & 13.20 & 10.70 & --\\
E92   & $V$ & -- & 11.90 & --\\
Z04   & $V$ & -- & 12.140 & --\\
F83   & $P$ & -- & 12.6 & --\\
L96   & $P$ & -- & 11.46$\pm$0.40 & --\\
K99   & $B$ & -- & 12.33 & --\\
U98   & $B$ & -- & 13.05 & --\\
 UKIDSS & $J$ & 11.872$\pm$0.001 & 10.977 & 10.753\\
       & $H$ & 11.737$\pm$0.001 & 10.942 & 10.906\\
       & $K$ & 11.498$\pm$0.001 & 10.451$\pm$0.001 & 10.456\\
 2MASS & $J$ & 11.796 & 10.868$\pm$0.021 & 10.602$\pm$0.021\\
      & $H$ & 11.579$\pm$0.030 & 10.588$\pm$0.016 & 10.433$\pm$0.018\\
      & $K$ & 11.223 & 10.392$\pm$0.018 & 10.400$\pm$0.018\\
{\it WISE} & $W1$ & 10.741$\pm$0.024 & 9.919$\pm$0.022 & 10.263$\pm$0.025\\
     & $W2$ & 10.541$\pm$0.021 & 9.699$\pm$0.020 & 10.307$\pm$0.023\\
     & $W3$ & 10.030$\pm$0.067 & 8.927$\pm$0.037 & 10.327$\pm$0.106\\
     & $W4$ & 8.558$\pm$0.285 & 7.617$\pm$0.124 & 8.964\\
\noalign{\smallskip}\hline\noalign{\smallskip}
\end{tabular}
\parbox{120mm}
{\baselineskip 3.6mm References: KW97 -- Kohoutek \& Wehmeyer
(1997), 
 JKP11 -- \cite{Jerzykiewicz+etal+2011}, HMUBV -- Homogeneous
Means in the UBV System 
(Mermilliod 1997), P06 -- \cite{Paunzen+etal+2006}, H61 --
\cite{Hoag+etal+1961}, F83 -- \cite{Fresneau+1983}, L96 -- Lasker
et al.\ (1996), K99 -- \cite{Kislyuk+etal+1999}, U98 --
 Lucas et al.\ (2008), E92 -- \cite{Egret+etal+1992}, Z04 --
\cite{Zacharias+etal+2004}, {\it UKIDSS} --, 2MASS --
\citealt{Cutri+etal+2003}, {\it WISE} -- Cutri et al.\ 2012}
\end{table*}

\begin{figure}
\centering
\includegraphics[width=8.9cm]{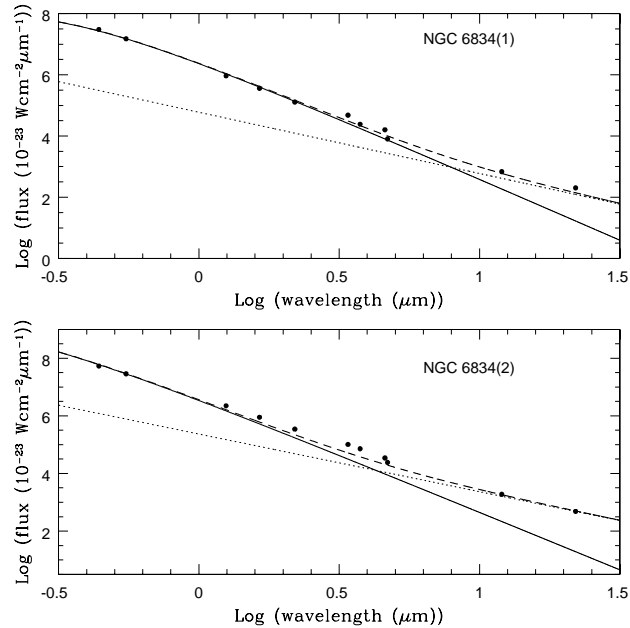}

\vspace{-3mm} \caption{\baselineskip 3.6mm SEDs of NGC~6834(1) and
NGC~6834(2). The dotted lines show the contribution from the
free-free emission, calculated as a function of wavelength. The
solid lines show the blackbody emission for temperatures
appropriate for the spectral types of the two stars. The dashed
lines show the total contributions from the blackbody and the
free-free components.\label{fig9}}
\end{figure}

\subsection{Spectral Energy Distribution}

The SED{s} of NGC~6834(1) and NGC~6834(2) are constructed using
the new observations in {the }$B$, $V$, $J$, $H$, $K$, $L'${ and}
$M'$ bands presented in this work (Tables~5 and 2 respectively),
and the {\it WISE} data in the four bands $W1$, $W2$, $W3$ and
$W4$ given in Table~\ref{tab5}. 
 The magnitudes are
extinction corrected for $E(B-V)$ = 0.73, using the relations in
\cite{Rieke+Lebofsky+1985}. The extinction corrected $BVJHKL'M'$
magnitudes are converted to flux using the zero-magnitude flux
values from \cite{Bessell+etal+1998} and {the }JAC/UKIRT
homepage\footnote{\it
http://www.jach.hawaii.edu/UKIRT/astronomy/calib/phot\_cal/conver.html}.
The {\it WISE} magnitudes {are} corrected for extinction and
converted to flux using the zero-magnitude flux values given in
the {\it WISE} explanatory supplement (Jarrett et al.\ 2011).

The SEDs of the Be stars NGC~6834(1) and NGC~6834(2) are shown in
Figure~\ref{fig9}. 
  The blackbody curves are generated
corresponding to the effective temperature of the stars, using the
values tabulated in \cite{Schmidt-Kaler+1982} for different
spectral types, and adopting B5V for NGC~6834(1) and B1-1.5V for
NGC~6834(2). The blackbody curves are normalized to the extinction
corrected $V$ band flux values of the two stars. It can be seen
from Figure~\ref{fig9} that the blackbody curves poorly fit the
SEDs in the infrared region, thereby indicating the presence of IR
excess in these stars. It has been {demonstrated} from various
studies that IR excess in Be stars is due to free-free emission
from free electrons in the dis{k} (\citealt{Gehrz+etal+1974}). In
\cite{Mathew+etal+2012b}, we used a simplified model to examine if
the free-free emission can account for the IR excess in the Be
star X Persei. In this work we use the same method to model the
excess flux seen in Figure~\ref{fig9}. The input parameters needed
for the model are the temperature, electron density, stellar
radius and the distance to the Be star. The radii of the Be stars
corresponding to their spectral types are taken from
\cite{Gehrz+etal+1974}. The cluster distance of 3100\,pc is used.
The electron density is a free parameter, whose value is optimized
from the fit to the data points. The calculated flux values of
free-free emission as a function of wavelength are shown by a
dotted line in Figure~\ref{fig9}. These flux values are calculated
for a temperature of {10\,000}\,K. We derive electron densities
($n_{\rm e}$) of 2$\times$10$^{11}$~cm$^{-3}$ for NGC~6834(1), and
1.5$\times$10$^{11}$~cm$^{-3}$ for NGC~6834(2). These values of
electron density agree with the values derived from theoretical
modeling by \cite{Silaj+etal+2010} for Be stars of similar
spectral types.

\section{DISCUSSION}
\subsection{On the Binarity of NGC~6834 (1)}

\cite{Mason+etal+2002}, from speckle interferometric studies,
suggested a possible companion to NGC~6834(1) at a separation of
10.35$\arcsec$ and a position angle of 74.7$\dg$. The system is
designated by Washington Speckle Interferometry (WSI) number
WSI\,12. They derived the $V$ magnitudes of the primary and
secondary as 11.7 and 12.1{,} respectively. The epoch of the
observation was 2001.738 (given as fractional B{e}sselian year). The
measurements of{ the} WSI\,12 system were repeated after {two}
years (Epoch = 2003.782) and the results are presented in
\cite{Mason+etal+2004}. The separation of the components and the
position angle are 10.96$\arcsec$ and 74.4$\dg${,} respectively.
The results from the second observations agree with the previous
estimates.

In the POSS~II $R$-band image (Fig.~\ref{fig1}), the star
immediate{ly beside} NGC~6834(1){,} located at RA = 19:52:05.70
and Dec = +29:24:36.4{,} is the possible companion reported by
\cite{Mason+etal+2002}. This is the only star at the distance of
10.35$\arcsec$ from the Be star. This star (denoted by
NGC~6834(1B) in this work) has been identified as a B-type giant
(B5III) by \cite{Turner+1976}. NGC~6834(1B) also appeared in the
photoelectric studies of \cite{Hoag+etal+1961} (numbered 7 while
Be star is numbered 16), whose magnitudes and colors are given in
Table~\ref{tab5}. From our new optical photometry it can be seen
that the $V$ mag of the Be star is 13.22 while that of the
proposed companion is 11.93. When compared to the Be star, the
B-type giant is brighter by 1.3\,mag in $V$, $\sim$1\,mag in {\it
JHK} and 0.2--0.5\,mag in the {\it WISE} bands. A tentative trend
of reversal in brightness, with the Be star brighter than {a
}B-type giant, is seen in{ the} W3 and W4 bands. This is due to
the fact that the IR excess from the{ disk of the} Be star starts
to dominate in this region (as shown in Fig.~\ref{fig9}) whereas
the blackbody{ emission from} NGC~6834(1B) keeps {decreasing}.

It can be seen from Table~\ref{tab5} that most of the photometric
measurements quote a $V$ mag of 13.2 for NGC~6834(1) and 11.9 for
NGC~6834(1B). If we compare these values with the speckle
measurements, NGC~6834(1B) match{es} very well whereas the Be star
was brighter by about 1\,mag during speckle measurements. The $V$
mag measurements by \cite{Jerzykiewicz+etal+2011} were also done
during the same period (April-October 2001) and they report a
value of 13.117 for the Be star. \cite{Paunzen+etal+2006} {also
}obtained a similar value ($V$ = 13.22) from their observations
conducted in August 2004. This presents a puzzle regarding the
brighter magnitude reported for the Be star in speckle
interferometric observations. The only measurement reported in the
literature which closely agrees with the value of
\cite{Mason+etal+2002} is from the NOMAD Catalog
(\citealt{Zacharias+etal+2004}), who reported a $V$ mag of 12.390.
This is a compilation of various observations done in the {US
}Naval Observatory and this particular measurement was taken from
the YB6 Catalog (USNO, unpublished). This difference between the
magnitudes of the Be star from photometry and speckle
interferometry (both of which are obtained during the same period)
hinders the discussion on the nature of the binary system. In the
future we may try to get radial velocity measurements of both
stars to look for {a }possible association between them.

\begin{figure}
\centering
\includegraphics[width=7.5cm]{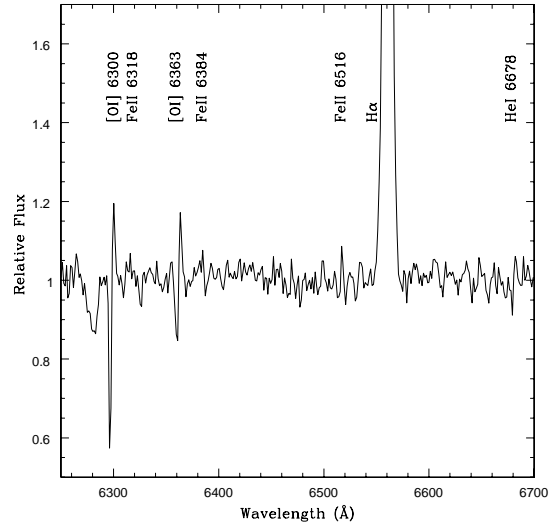}

\vspace{-3mm} \caption{\baselineskip 3.6mm Optical spectra
(6250--6700~\AA) of NGC~6834(1) observed on 2007-11-14 using the
OMR spectrograph mounted on VBT showing [O{\sc i}] 6300,
6363\,\AA~emission lines.\label{fig10}}
\end{figure}

Another interesting observation worth reporting is the detection
of the [O{\sc i}] 6300 and 6363\,\AA\ emission lines in the
spectra of NGC~6834(1), obtained at a resolution of 4\,\AA\ from{
the} 2.34-m VBT, Kavalur (Fig.~\ref{fig10}). 
 The
[O{\sc i}] 6300, 6363\,\AA\ lines show {a }P-Cygni feature, which
is indicative of the mass-loss process in the Be star
(Fig.~\ref{fig10}). If the lines are formed in the outflow, the
outflow velocity can be calculated from the difference between the
absorption and emission components of [O{\sc i}] lines. The
estimated value of 124 km~s$^{-1}$ (for both the lines) closely
matches the values estimated for young stellar objects (YSO{s};
\citealt{Acke+etal+2005}). The chance of NGC~6834(1) being a YSO
is feeble since such a pre-main sequence star is generally not
found in{ an} 80 Myr old cluster. {In addition}, from the optical
CMD (Fig.~\ref{fig3}) we found that the star is close to the
evolved part of the main sequence. There are no reported cases of
the presence of [O{\sc i}] lines (especially with P-Cygni nature)
in classical Be stars and the significance of this intriguing
observation needs to be {better }understood.

Since NGC~6834 (1B) {is separated from the Be star by} 10.35$\arcsec$, the chance of the giant influencing the Be star is minimal.
{In addition}, the reason for outflow (deduced from
[O{\sc i}]lines) in NGC~6834(1) needs to be analy{z}ed from{ a} fresh set
of observations. The role of a hidden companion
trig{g}ering this outflow has to be explored even though
we do not identify any high energy emission associated with the
system.

\subsection{NGC~6834 (2) - a $\gamma$ Cas-like Variable?}

$\gamma$ Cas displays moderately high X-ray luminosity
(10$^{32}-10^{33}$~erg~s$^{-1}$), which is an order of magnitude
higher than {that} seen in massive stars of a similar spectral
type. {In addition}, the X-ray emission is characterized by a hot
thermal component of $T$$\sim$140--165~MK whereas the plasma
temperature for soft X-ray emission in other massive stars is
identified to be 6~MK (\citealt{LopesdeOliveira+Motch+2011}). The
X-ray spectrum of $\gamma$ Cas is similar to that of cataclysmic
variables, which support{s} the idea that accretion onto the white
dwarf is responsible for the high energy emission. Alternatively,
the high energy component in X-ray emission can also be due to the
magnetic reconnection at the interface between the photosphere and
the inner part of the dis{k} (\citealt{Smith+Robinson+2003}).
$\gamma$ Cas-like variables are Be stars {that} are
characteri{z}ed by{ an} unusually hard thermal X-ray spectrum,
variable on short time scales, and an early B-type optical
spectrum with a dense circumstellar disk.

\cite{Jerzykiewicz+etal+2011} found that NGC~6834(2) exhibits
short-term variability in $V$ and $I_C$ bands, with amplitudes of
0.23 and 0.22 mag respectively. This variability in optical bands
along with the strong H$\alpha$ emission prompted
\cite{Jerzykiewicz+etal+2011} to include NGC~6834(2) in the class of
$\gamma$ Cas-like variables. However, they have not mentioned
any X-ray studies of this object.

$\gamma$ Cas-like variables are characterized in the optical and
infrared by means of the variability in their light curves, strong
H$\alpha$ emission and IR excess. The strong H$\alpha$ emission
along with the infrared excess is indicative of a dense
circumstellar disk (\citealt{LopesdeOliveira+Motch+2011}). Excess
emission {in the} near- and mid-infrared region is clearly
{evident} in the SED of NGC~6834(2) (Fig.~\ref{fig9}). A strong
H$\alpha$ emission (EW = 42\,\AA) is noticed in NGC~6834(2), which
is found to be variable, fading by 7\,\AA\ over a period of {four}
months (see Table~\ref{tab3}). From the archival data we found
that NGC~6834(2) showed long-term variability of 0.7--1.1\,mag in
$V$ (see Table~\ref{tab5}). \cite{Kohoutek+Wehmeyer+1997} reported
the star to be brighter by 1.7~mag in the $V$-band ($V$ = 10.7),
which is indicative of flaring activity during that period. {In
addition}, \cite{Jerzykiewicz+etal+2011} identified short-term
variability of 0.2~mag. Hence, this fulfills the criteria required
in optical and infrared for the Be star to belong to the class of
$\gamma$ Cas-like variables. However, in addition to the above
requirements, moderately high X-ray emission is a necessary
criteri{on} to classify the object as a $\gamma$ Cas-like
variable. Since we {have not found} any reports of X-ray (or high
energy) emissions from this object, we are not certain about the
classification.

\begin{acknowledgements}
We thank the anonymous reviewer for helpful comments.
The UKIRT is operated by the Joint Astronomy Centre on
behalf of the Science and Technology Facilities Council (STFC) of the UK.
We thank the UKIRT service program for obtaining the IR observations.
We would like to thank the support astronomers in HCT and VBT.
The research work at{ the} Physical Research Laboratory is funded by the Department
of Space, Government of India. This research has made use of the VizieR
catalogue access tool, CDS, Strasbourg, France. The original description of
the VizieR service was published in A\&AS 143, 23.
\end{acknowledgements}

\end{document}